\documentclass[12pt,letterpaper]{article}
\usepackage{amsmath,amsfonts,amsthm,amssymb}

\begin{document}

\newcounter{rom} 

\title{(Compression Bases in Unital Groups}
\author{David J. Foulis{\footnote{Emeritus Professor, 
University of Massachusetts; foulis@math.umass.edu;
1 Sutton Court, Amherst, MA 01002, USA.}}} 
\date{}
\maketitle

\begin{abstract}
We introduce and launch a study of compression bases in unital 
groups. The family of all compressions on a compressible group 
and the family of all direct compressions on a unital group 
are examples of compression bases. In this article we show that 
the properties of the compatibility relation in a compressible 
group generalize to unital groups with compression bases. 

\end{abstract}

\medskip

\noindent {\bf AMS Classification:} Primary 06F20.  
Secondary 47A20.

\medskip

\noindent {\bf Key Words and Phrases:} normal sub-effect 
algebra, compatibility, unital group, compression, 
compressible group, compression base.  

\section{Normal Sub-Effect Algebras}  % Section 1

If $E$ is an effect algebra \cite{FB}, then a {\em Mackey 
decomposition} in $E$ of the ordered pair $(e,f)\in E\times 
E$ is an ordered triple $(e\sb{1},f\sb{1},d)\in E\times 
E\times E$ such that $e\sb{1}\perp f\sb{1}$, $(e\sb{1}
\oplus f\sb{1})\perp d$, $e=e\sb{1}\oplus d$, and $f
=f\sb{1}\oplus d$. If there exists a Mackey decomposition 
in $E$ of $(e,f)\in E\times E$, then $e$ and $f$ are said to 
be {\em Mackey compatible} in $E$. 

\medskip

\noindent{\bf 1.1 Definition} Let $P$ be a sub-effect 
algebra of the effect algebra $E$ \cite[Definition 2.6]{FB}. 
Then $P$ is a {\em normal} sub-effect algebra of $E$ iff, 
for all $e,f\in P$, if $(e\sb{1},f\sb{1},d)\in E\times E
\times E$ is a Mackey decomposition in $E$ of $(e,f)$, 
then $d\in P$.

\medskip

Suppose that $E$ is an effect algebra, $P$ is a sub-effect 
algebra of $E$, $e,f\in P$, and $(e\sb{1},f\sb{1},d)\in 
E\times E\times E$ is a Mackey decomposition of $(e,f)$ 
in $E$. Then $e$ and $f$ are Mackey compatible in $E$, but 
not necessarily in $P$. However, if $P$ is a normal 
sub-effect algebra of $E$, then $d\in P$ and, since 
$e\sb{1}\oplus d=e$, $f\sb{1}\oplus d=f$, and $d,e,f\in 
P$, it follows that $e\sb{1},f\sb{1}\in P$, whence 
$(e\sb{1},f\sb{1},d)\in P\times P\times P$ is a Mackey 
decomposition in $P$ of $(e,f)$.  Therefore, {\em 
if $P$ is a normal sub-effect algebra of $E$ and $e,f
\in P$, then $e$ and $f$ are Mackey compatible in $E$ 
iff $e$ and $f$ are Mackey compatible in $P$.}

\medskip

\noindent{\bf 1.2 Example} The center of an effect 
algebra $E$ \cite{GFP} is a normal sub-effect algebra 
of $E$.

\medskip

Recall that $G$ is a {\em unital group} with {\em unit} 
$u$ and {\em unit interval $E$} iff $G$ is a directed 
partially ordered abelian group \cite{Good}, such that 
$u\in G\sp{+} :=\{g\in G\mid 0\leq g\}$, $E :=\{e\in G
\mid 0\leq e\leq u\}$, and every element $g\in G\sp{+}$ 
can be written as $g=\sum\sb{i=1}\sp{n}e\sb{i}$ with 
$e\sb{i}\in E$ for $i=1,2,...,n$ \cite[p. 436]{FCG}. 
(The symbol $:=$ means ``equals by definition.") 

Suppose that $G$ is a unital group with unit $u$ and 
unit interval $E$. Then $E$ is an effect algebra with 
unit $u$ under the partially defined binary operation 
$\oplus$ obtained be restriction of $+$ on $G$ to $E$ 
\cite{BF}.  We note that a sub-effect algebra $P$ of 
$E$ is a normal sub-effect algebra of $E$ iff, for all 
$e,f,d\in E$ with $e+f+d\leq u$, we have $e+d,f+d\in P
\Rightarrow d\in P$.

\medskip

\noindent{\bf 1.3 Example} Let $\mathfrak H$ be a Hilbert 
space.  Then the additive abelian group $\mathbb G$ of all 
bounded self-adjoint operators on $\mathfrak H$, partially 
ordered in the usual way, is a unital group with unit 
$\mathbf 1$. The unit interval $\mathbb E$ in $\mathbb G$ 
is the standard effect algebra of all effect operators on 
$\mathfrak H$, and the orthomodular lattice $\mathbb P$ of 
all projection operators on $\mathfrak H$ is a normal 
sub-effect algebra of $\mathbb E$.\hspace{\fill}$\square$ 

\section{Retractions and Compressions} %Section 2
   
Let $G$ be a unital group with unit $u$ and unit interval 
$E$. A {\em retraction} on $G$ with {\em focus $p$} is 
defined to be an order-preserving group endomorphism $J\colon 
G\to G$ with $p=J(u)\in E$ such that, for all $e\in E$, 
$e\leq p\Rightarrow J(e)=e$. A retraction $J$ on $G$ with 
focus $p$ is called a {\em compression} on $G$ iff $J(e)=
0\Rightarrow e\leq u-p$ holds for all $e\in E$ \cite{FC}.

The unital group $G$ always admits at least two compressions, 
namely the zero mapping, $g\mapsto 0$ for all $g\in G$ and the 
identity mapping $g\mapsto g$ for all $g\in G$. Conversely, 
the only retraction on $G$ with focus $0$ is the zero mapping, 
and the only retraction on $G$ with focus $u$ is the identity 
mapping. Suppose $J$ is a retraction with focus $p$ on $G$. 
Then, $J$ is idempotent (i.e., $J\circ J=J$) and $J(p)=p$. 
Also, for all $e\in E$, $e\leq u-p\Rightarrow J(e)=0$ and, 
if $J$ is a compression, then $e\leq u-p\Leftrightarrow J(e)
=0$ \cite{FC}.

\medskip

\noindent{\bf 2.1 Lemma} {\em Let $G$ be a unital group 
with unit $u$ and unit interval $E$. Suppose that $J$ is 
a compression on $G$ with focus $p$, and $J\,'$ is a 
retraction on $G$ with focus $u-p$. Then, for all 
$g\in G\sp{+}$, $J(g)=0\Leftrightarrow J\,'(g)=g$.}

\medskip

\noindent{\em Proof} Let $e\in E$. As $0\leq e\leq u$, we 
have $0\leq J\,'(e)\leq J\,'(u)=u-p$, whence $J(J\,'(e))=0$. 
Since $E$ generates $G$ as a group and $J\circ J\,'$ is an 
endomorphism on $G$, we have $J(J\,'(g))=0$ for all $g\in G$. 
As $J$ is a compression with focus $p$, it follows that $J(e)
=0\Rightarrow e\leq u-p\Rightarrow J\,'(e)=e$. Now let $g\in 
G\sp{+}$ and write $g=\sum\sb{i=1}\sp{n}e\sb{i}$ with $e\sb{i}
\in E$ for $i=1,2,...,n$. If $J(g)=0$, then $\sum\sb{i=1}\sp{n}
J(e\sb{i})=0$ and, since $0\leq J(e\sb{i})$ for $i=1,2,...,n$, 
it follows that $J(e\sb{i})=0$ for $i=1,2,...,n$, whence $J\,'
(e\sb{i})=e\sb{i}$ for $i=1,2,...,n$, and therefore  
$J\,'(g)=g$. Conversely, if $J\,'(g)=g$, then $J(g)=J
(J\,'(g))=0$.\hspace{\fill}$\square$

\medskip

A {\em compressible group} is defined to be a unital group 
$G$ such that (1) every retraction on $G$ is uniquely 
determined by its focus, and (2) if $J$ is a retraction 
on $G$, there exists a retraction $J\,'$ on $G$ such that, 
for all $g\in G\sp{+}$, $J(g)=0\Leftrightarrow J\,'(g)=g$ 
and $J\,'(g)=0\Leftrightarrow J(g)=g$ \cite[Definition 3.3]
{FC}. If $G$ is a compressible group, then an element $p\in G$ 
is called a {\em projection} iff it is the focus of a retraction 
on $G$.  Suppose that $G$ is a compressible group and $P$ 
is the set of all projections in $G$.  Then every retraction 
on $G$ is a compression, and if $p\in P$, then the unique 
retraction (hence compression) on $G$ with focus $p$ is 
denoted by $J\sb{p}$. The set $P$ is a sub-effect algebra 
of $E$ and, in its own right, it forms an orthomodular 
poset (OMP) \cite[Corollary 5.2 (iii)]{FCG}.

\medskip

\noindent{\bf 2.2 Example}  Let $A$ be a unital 
C$\sp{\ast}$-algebra and let $G$ be the additive group 
of all self-adjoint elements in $A$.  Then $G$ is a unital 
group with unit $1$ and positive cone $G\sp{+}=\{aa\sp{\ast}
\mid a\in A\}$.  The unital group $G$ is a compressible 
group with $P=\{p\in G\mid p=p\sp{2}\}$ and, if $p\in P$, 
then $J\sb{p}(g)=pgp$ for all $g\in G$ \cite{FC}.\hspace
{\fill}$\square$

\medskip

\noindent{\bf 2.3 Theorem} {\em Let $G$ be a compressible 
group with unit $u$ and unit interval $E$.  Then:} (i) 
{\em $P$ is a normal sub-effect algebra of $E$.} (ii) 
{\em If $p,q,r\in P$ with $p+q+r\leq u$, then $J\sb
{p+r}\circ J\sb{q+r}=J\sb{r}$.}

\medskip

\noindent{\em Proof} (i) By \cite[Corollary 5.2 (ii)]{FCG}, 
$P$ is a sub-effect algebra of $E$. Suppose $e,f,d\in E$, 
$e+f+d\leq u$, $e+d\in P$, $f+d\in P$, and define $J :=
J\sb{e+d}\circ J\sb{f+d}$. Then $J\colon G\to G$ is an 
order-preserving endomorphism and $J(u)=J\sb{e+d}
(J\sb{f+d}(u))=J\sb{e+d}(f+d)=J\sb{e+d}(f)+J\sb{e+d}(d)$. 
But, $e+f+d\leq u$, so $f\leq u-(e+d)$, and $d\leq e+d$, 
whence $J(u)=0+d=d$. Suppose $h\in E$ with $h\leq d$. Then 
$h\leq e+d,f+d$, and it follows that $J(h)=J\sb{e+d}
(J\sb{f+d}(h))=J\sb{e+d}(h)=h$. Therefore $J$ is a 
retraction with focus $d$, so $d\in P$.
 
(ii) If $p,q,r\in P$ and $p+q+r\leq u$, then by the proof 
of (i) above with $e$ replaced by $p$, $f$ replaced by $q$, 
and $d$ replaced by $r$, we have $J\sb{p+r}\circ J\sb{q+r}
=J\sb{r}$.\hspace{\fill}$\square$
   
\section{Compression Bases} %Section 3

By Theorem 2.3, the notion of a ``compression base," as per 
the following definition, generalizes the family $(J\sb{p})
\sb{p\in P}$ of compressions in a compressible group.

\medskip

\noindent{\bf 3.1 Definition} Let $G$ be a unital group with 
unit interval $E$. A family $(J\sb{p})\sb{p\in P}$ of 
compressions on $G$, indexed by a normal sub-effect algebra 
$P$ of $E$, is called a {\em compression base} for $G$ iff 
(i) each $p\in P$ is the focus of the corresponding compression 
$J\sb{p}$, and (ii) if $p,q,r\in P$ and $p+q+r\leq u$, then 
$J\sb{p+r}\circ J\sb{q+r}=J\sb{r}$.

\medskip

The conditions for a unital group to be a compressible group 
are quite strong and they rule out many otherwise interesting 
unital groups. On the other hand, the notion of a unital group 
$G$ with a specified compression base $(J\sb{p})\sb{p\in P}$ 
is very general, yet most of the salient properties of projections 
and compressions for a compressible group generalize, mutatis 
mutandis, to the elements $p\in P$ and to the compressions 
$J\sb{p}$ in the compression base for $G$.

\medskip

\noindent{\bf 3.2 Example} A retraction $J$ on the unital group 
$G$ is {\em direct} iff $J(g)\leq g$ for all $g\in G\sp{+}$ 
\cite[Definition 2.6]{FC}. For instance, the zero mapping $g
\mapsto 0$ and the identity mapping $g\mapsto g$ for all $g\in 
G$ are direct compressions on $G$. Let $P$ be the set of all 
foci of direct retractions on $G$. Then $P$ is a sub-effect 
algebra of the center of $E$. Also, if $p\in P$, there is a 
unique retraction $J\sb{p}$ on $G$ with focus $p$, and $J\sb{p}$ 
is a compression. Furthermore, the family $(J\sb{p})\sb{p\in P}$ 
is a compression base for $G$.\hspace{\fill}$\square$

\medskip

\noindent{\bf 3.3 Standing Assumption} In the sequel, {\em 
we assume that $G$ is a unital group with unit $u$ and unit 
interval $E$ and that $(J\sb{p})\sb{p\in P}$ is a compression 
base for $G$.} 

\medskip

\noindent{\bf 3.4 Theorem} {\em $P$ is an orthomodular 
poset and, if $p\in P$ and $g\in G\sp{+}$, then $J\sb{p}
(g)=0 \Leftrightarrow J\sb{u-p}(g)=g$.}

\medskip

\noindent{\em Proof} By \cite[Lemma 2.3 (iv)]{FC}, every 
element in $P$ is a principal, hence a sharp, element of 
$E$. Therefore, $P$ is an OMP. That $J\sb{p}(g)=0
\Leftrightarrow J\sb{u-p}(g)=g$ for $p\in P$ and $g\in 
G\sp{+}$ follows from Lemma 2.1.\hspace{\fill}$\square$

\medskip

\noindent{\bf 3.5 Lemma} {\em If $p,q\in P$, then the 
following conditions are mutually equivalent:} 
(i) $q\leq p$. (ii) $J\sb{p}\circ J\sb{q}=J\sb{q}$. 
(iii) $J\sb{p}(q)=q$. (iv) $J\sb{q}\circ J\sb{p}=
J\sb{q}$. (v) $J\sb{q}(p)=q$.

\medskip

\noindent{\em Proof} (i) $\Rightarrow$ (ii). Assume (i). 
Then $p-q\in P$ and $(p-q)+0+q=p\leq u$, hence, by Definition 
3.1 (ii), $J\sb{(p-q)+q}\circ J\sb{0+q}=J\sb{q}$, i.e., 
$J\sb{p}\circ J\sb{q}=J\sb{q}$.

(ii) $\Rightarrow$ (iii).  Assume (ii).  Then $J\sb{p}(q)
=J\sb{p}(J\sb{q}(u))=J\sb{q}(u)=q$.

(iii) $\Rightarrow$ (iv). Assume (iii). Then $q=J\sb{p}(q)
\leq p$. Therefore $p-q\in P$ and $0+(p-q)+q=p\leq u$; hence, 
by Definition 3.1 (ii), $J\sb{0+q}\circ J\sb{(p-q)+q}=
J\sb{q}$, i.e., $J\sb{q}\circ J\sb{p}=J\sb{q}$.

(iv) $\Rightarrow$ (v). Assume (iv). Then $q=J\sb{q}(u)=
J\sb{q}(J\sb{p}(u))=J\sb{q}(p)$. 

(v) $\Rightarrow$ (i). Assume (v). Then $J\sb{q}(u-p)=
q-q=0$, so $u-p=J\sb{u-q}(u-p)=J\sb{u-q}(u)-J\sb{u-q}(p)
=u-q-J\sb{u-q}(p)$, i.e., $q+J\sb{u-q}(p)=p$.  But $0
\leq J\sb{u-q}(p)$, so $q\leq p$.\hspace{\fill}$\square$

\section{Compatibility} %Section 4

We maintain our standing assumption that $(J\sb{p})
\sb{p\in P}$ is a compression base for the unital group 
$G$ with unit $u$ and unit interval $E$. The notion 
of compatibility in a compressible group \cite
[Definition 4.1]{FCG} carries over, as follows, to 
$G$. 

\medskip

\noindent{\bf 4.1 Definition} If $p\in P$, we define 
$C(p) :=\{g\in G\mid g=J\sb{p}(g)+J\sb{u-p}(g)\}$. 
If $g\in C(p)$, we say that $g$ is {\em compatible} 
with $p\in P$. For $p,q\in P$, we often write the 
condition $q\in C(p)$ in the alternative form $qCp$.

\medskip

We devote the remainder of this article to showing that 
{\em the fundamental properties of compatibility in a
compressible group generalize to a unital group with a 
compression base.}

\medskip

\noindent{\bf 4.2 Lemma} {\em Let $p\in P$ and $g\in G$. 
Then $J\sb{p}(g)\leq g\Rightarrow g\in C(p)$, and  
$0\leq g\in C(p)\Rightarrow J\sb{p}(g)\leq g$.} 
 
\medskip

\noindent{\em Proof} Suppose $J\sb{p}(g)\leq g$.  Then 
$0\leq g-J\sb{p}(g)$ and $J\sb{p}(g-J\sb{p}(g))=J\sb{p}(g)
-J\sb{p}(g)=0$, whence $g-J\sb{g}(g)=J\sb{u-p}(g-J\sb{p}(g))
=J\sb{u-p}(g)-0=J\sb{u-p}(g)$, i.e., $g=J\sb{p}(g)+J\sb{u-p}(g)$, 
and therefore, $g\in C(p)$.  Conversely, if $0\leq g\in C(p)$, 
then $0\leq J\sb{u-g}(g)$, whence $J\sb{p}(g)\leq J\sb{p}(g)
+J\sb{u-p}(g)=g$.\hspace{\fill}$\square$

\medskip

\noindent{\bf 4.3 Theorem} {\em Let $p,q\in P$.  Then 
the following conditions are mutually equivalent:}
(i) $J\sb{p}\circ J\sb{q}=J\sb{q}\circ J\sb{p}$. (ii) 
$J\sb{p}(q)=J\sb{q}(p)$. (iii) $J\sb{p}(q)\leq q$. (iv) 
$p$ {\em is Mackey compatible with $q$ in $E$.} (v) $p$ 
{\em is Mackey compatible with $q$ in $P$.} (vi) $\exists 
r\in P, J\sb{p}\circ J\sb{q}=J\sb{r}$. (vii) $J\sb{p}(q)\in 
P$. (viii) $qCp$

\medskip

\noindent{\em Proof} (i) $\Rightarrow$ (ii). If (i) holds, 
then $J\sb{p}(q)=J\sb{p}(J\sb{q}(u))=J\sb{q}(J\sb{p}(u))=
J\sb{q}(p)$.

(ii) $\Rightarrow$ (iii). If (ii) holds, then $J\sb{p}(q)
=J\sb{q}(p)\leq q$.

(iii) $\Rightarrow$ (iv). Let $r :=J\sb{p}(q)$ and assume that 
$r\leq q$. Then $0\leq r\leq p,q$, whence $e :=p-r\in E$ and 
$f :=q-r\in E$ with $e+r=p$ and $f+r=q$. As $J\sb{p}(f)=
J\sb{p}(q-r)=r-r=0$, we have $f\leq u-p$, whence $e+f+r=
f+p\leq u$, and it follows the $p$ is Mackey compatible with 
$q$ in $E$.

(iv) $\Rightarrow$ (v). As $P$ is a normal sub-effect algebra 
of $E$, we have (iv) $\Rightarrow$ (v).

(v) $\Rightarrow$ (vi). If (v) holds, there exist $e,f,r\in 
P$ with $e+f+r\leq u$, $p=e+r$ and $q=f+r$. Therefore, by 
Definition 3.1 (ii), $J\sb{p}\circ J\sb{q}=J\sb{e+r}\circ 
J\sb{f+r}=J\sb{r}$.

(vi) $\Rightarrow$ (vii). Suppose that $r\in P$ and $J\sb{p}
\circ J\sb{q}=J\sb{r}$.  Then $J\sb{p}(q)=J\sb{p}(J\sb{q}(u))
=J\sb{r}(u)=r\in P$.

(vii) $\Rightarrow$ (viii) Assume (vii) and let $r :=J\sb{p}(q)
\in P$. Then $J\sb{r}(q)\leq r\leq p$, so $0\leq r-J\sb{r}(q)$. 
Thus, by Lemma 3.5, $r-J\sb{r}(q)=r-(J\sb{r}\circ J\sb{p})(q)=
r-J\sb{r}(J\sb{p}(q))=r-J\sb{r}(r)=r-r=0$, i.e., $r=J\sb{r}
(q)$. Therefore, $J\sb{r}(u-q)=r-r=0$, so $u-q\leq u-r$, i.e., 
$r\leq q$, and it follows from Lemma 4.2 that $pCq$.

(viii) $\Rightarrow$ (i). Assume that $qCp$.  Then, by Lemma 
4.2, $J\sb{p}(q)\leq q$, so (iii) holds. We have already shown 
that (iii) $\Rightarrow$ (iv) $\Rightarrow (v)$, so there 
exist $e,f,r\in P$ with $e+f+r\leq u$, $p=e+r$, and $q=f+r$. 
Therefore, by Definition 3.1 (ii), $J\sb{p}\circ J\sb{q}=
J\sb{e+r}\circ J\sb{f+r}=J\sb{r}=J\sb{f+r}\circ J\sb{e+r}=
J\sb{q}\circ J\sb{p}$.\hspace{\fill}$\square$

\medskip

Because conditions (i), (ii), (iv), and (v) in Theorem 4.3 
are symmetric in $p$ and $q$, so are conditions (iii), 
(vi), (vii), and (viii). In particular, for $p,q\in P$, we 
have $pCq\Leftrightarrow qCp$.

\medskip

\noindent{\bf 4.4 Corollary} {\em Let $p,q\in P$ and 
suppose that $pCq$. Then $J\sb{q}(p)=J\sb{p}(q)=p\wedge q$ 
is the greatest lower bound of $p$ and $q$ both in $E$ and 
in $P$, and $J\sb{p}\circ J\sb{q}=J\sb{q}\circ J\sb{p}=
J\sb{p\wedge q}$.}

\medskip

\noindent{\em Proof} Suppose that $p,q\in P$ and $pCq$. 
By Theorem 4.3, there exists $r\in P$ with $J\sb{p}
\circ J\sb{q}=J\sb{q}\circ J\sb{p}=J\sb{r}$. Thus, 
$r=J\sb{p}(J\sb{q}(u))=J\sb{p}(q)=J\sb{q}(p)\leq p,q$.
If $e\in E$ with $e\leq p,q$, then $e=J\sb{p}(J\sb{q}
(e))=J\sb{r}(e)\leq r$, so $r$ is the greatest lower 
bound of $p$ and $q$ in $E$, hence also in $P$.\hspace
{\fill}$\square$

\medskip

\noindent{\bf 4.5 Theorem} {\em Let $v\in P$ and 
define $H :=J\sb{v}(G)$, $E\sb{H} :=\{e\in E\mid e\leq v\}$, 
and $P\sb{H} :=\{q\in P\mid q\leq v\}$. For each $q\in 
P\sb{H}$, let $J\sp{H}\sb{q}$ be the restriction of 
$J\sb{q}$ to $H$. Then:} (i) {\em With the induced partial 
order, $H=\{h\in G\mid h=J\sb{v}(h)\}$ is a unital group with 
unit $v$ and unit interval $H\cap E=E\sb{H}$.} (ii) {\em 
$H\cap P=P\sb{H}$, and if $q\in P\sb{H}$, then $J\sp{H}\sb{q}$ 
is a compression on $H$}. (iii) {\em $P\sb{H}$ is a normal 
sub-effect algebra of $E\sb{H}$.} (iv) $(J\sp{H}\sb{q})\sb
{q\in P\sb{H}}$ {\em is a compression base for $H$.}

\medskip

\noindent{\em Proof} (i) By \cite[Lemma 2.4]{FCG}, $H$ is 
a unital group with unit $v$ and unit interval $H\cap E$. 
As $J\sb{v}$ is idempotent, $H=\{h\in G\mid h=J\sb{v}(h)\}$. 
Thus, for $e\in E$, $e\leq v\Leftrightarrow e=J\sb{v}(e)
\Leftrightarrow e\in H$, whence $H\cap E=\{e\in E\mid 
e\leq v\}$. 

(ii) As $P\subseteq E$, we have $H\cap P=P\sb{H}$. If 
$h\in H$ and $q\in P\sb{H}$, then by Lemma 3.5, $J\sb{q}
(h)=J\sb{v}(J\sb{q}(h))\in H$. Therefore $J\sp{H}\sb{q}
\colon H\to H$ is an order-preserving group endomorphism, 
and by Lemma 3.5 again, $J\sp{H}\sb{q}(v)=J\sb{q}(v)=q$. 
Also, if $e\in E\sb{H}$ with $e\leq q$, then $J\sp{H}\sb{q}
(e)=J\sb{q}(e)=e$, so $J\sp{H}\sb{q}$ is a retraction on $H$. 
Suppose $e\in E\sb{H}$ and $J\sp{H}\sb{q}(e)=0$. Then $e\leq 
u-q$, so $e+q\leq u$. By \cite[Lemma 2.3 (iv)]{FC}, $v$ is a 
principal element of $E$, hence, since $0\leq e,q\leq v$, it 
follows that $e+q\leq v$, i.e., $e\leq v-q$.  Therefore, 
$J\sp{H}\sb{q}$ is a compression on $H$.

(iii) Suppose $e,f,d\in E\sb{H}$, $e+f+d\leq v$, and $e+d,
f+d\in P\sb{H}$. Then $e,f,d\in E$, $e+f+d\leq v\leq u$, 
and $e+d,f+d\in P$. As $P$ is a normal sub-effect algebra of 
$E$, it follows that $d\in P$. But $d\leq v$, so $d\in P
\sb{H}$.

(iv) Suppose $s,t,r\in P\sb{H}$ with $s+t+r\leq v$. Then 
$s,t,r\in P$ with $s+t+r\leq u$, whence $J\sb{s+r}\circ 
J\sb{t+r}=J\sb{r}$, and it follows that $J\sp{H}\sb{s+r}\circ 
J\sp{H}\sb{t+r}=J\sp{H}\sb{r}$.\hspace{\fill}$\square$

\medskip

\noindent{\bf 4.6 Theorem} {\em Let $v\in P$ and define 
$C :=C(v)$. For each $s\in C\cap P$, let $J\sp{C}\sb{s}$ be 
the restriction of $J\sb{s}$ to $C$. Then:} (i) {\em With the 
induced partial order, $C$ is a unital group with unit $u$ and 
unit interval $C\cap E=\{e+f\mid e,f\in E, e\leq v, f\leq 
u-v\}$.} (ii) {\em If $s\in C\cap P$, then $J\sp{C}\sb{s}$ is 
a compression on $C$}. (iii) {\em $C\cap P$ is a normal 
sub-effect algebra of $C\cap E$.} (iv) $(J\sp{C}\sb{s})\sb
{s\in C\cap P}$ {\em is a compression base for $C$.} 

\medskip

\noindent{\em Proof} Part (i) follows from \cite[Lemma 4.2 (iv)]
{FCG}, part (iii) is obvious, and part (iv) is easily confirmed 
once part (ii) is proved. To prove part (ii), assume that $g\in 
C=C(v)$ and $s\in P\cap C$. Then, by Lemma 3.5, $J\sp{C}\sb{s}(g)
=J\sb{s}(J\sb{v}(g)+J\sb{u-v}(g))=J\sb{s}(J\sb{v}(g))+J\sb{s}
(J\sb{u-v}(g))=J\sb{v}(J\sb{s}(g))+J\sb{u-v}(J\sb{s}(g))$, so 
$J\sp{C}\sb{s}(g)=J\sb{s}(g)\in C(v)=C$. Therefore $J\sp{C}\sb{s}
\colon C\to C$ is an order-preserving group endomorphism, hence 
it is obviously a compression on $C$.\hspace{\fill}$\square$    

\medskip

\noindent{\bf 4.7 Definition} If $C$ and $W$ are unital groups 
with units $u$ and $w$, respectively, and if  $(J\sp{C}\sb{q})
\sb{q\in Q}$ and $(J\sp{W}\sb{t})\sb{t\in T}$ are compression bases 
in $C$ and $W$, respectively, then an order-preserving group homomorphism 
$\phi\colon C\to W$ is called a {\em morphism of unital groups with 
compression bases} iff $\phi(u)=w$, $\phi(Q)\subseteq T$, and $J\sp{W}\sb{\phi(q)}\circ\phi=\phi\circ J\sp{C}\sb{q}$ for all 
$q\in Q$.

\medskip

We omit the straightforward proof of the following theorem.

\medskip

\noindent{\bf 4.7 Theorem} {\em Suppose $v\in P$ and define $H 
:=J\sb{v}(G)$, $K :=J\sb{u-v}(G)$, and $C :=C(v)$. Organize $H$, 
$K$, and $C$ into unital groups with compression bases $(J\sp{H}
\sb{q})\sb{q\in P\sb{H}}$, $(J\sp{K}\sb{r})\sb{r\in P\sb{K}}$, 
and $(J\sp{C}\sb{s})\sb{s\in C\cap P}$, respectively, as in 
Theorems 4.5 and 4.6. Let $\eta$ be the restriction to $C$ of 
$J\sb{v}$ and let $\kappa$ be the restriction to $C$ of $J\sb{u-v}$. 
Then $\eta\colon C\to H$ and $\kappa\colon C\to K$ are surjective 
morphisms of unital groups with compression bases and, in the 
category of unital groups with compression bases, $\eta$ and 
$\kappa$ provide a representation of $C$ as a direct product 
of $H$ and $K$.}

\medskip

In subsequent papers we shall prove that all of the major results 
in \cite{FCG,FC,FCGGC,Spec} can be generalized to unital groups 
with compression bases.

\end{document}